\documentclass[twocolumn,aps,prb,showpacs,preprintnumbers]{revtex4}%
\usepackage{mathrsfs}
\usepackage{amsmath}
\usepackage{amsfonts}
\usepackage{amssymb}
\usepackage{amsthm}
\usepackage{graphicx}
\usepackage{color}
\usepackage{graphics}
\usepackage{epsfig}
\usepackage{subfigure}%
\providecommand{\U}[1]{\protect\rule{.1in}{.1in}}
\begin{document}
\title{Spin Wave Emission in Field-Driven Domain Wall Motion}
\author{X.S. Wang}
\author{X.R. Wang}
\affiliation{Physics Department, The Hong Kong University of
Science and Technology, Clear Water Bay, Kowloon, Hong Kong}
\affiliation{HKUST Shenzhen Research Institute, Shenzhen 518057, China}

\begin{abstract}
A domain wall (DW) in a nanowire can propagate under a longitudinal
magnetic field by emitting spin waves (SWs). We numerically
investigated the properties of SWs emitted by the DW motion, such
as frequency and wavenumber, and their relation with the DW motion.
For a wire with a low transverse anisotropy and in a field above
a critical value, a DW emits SWs to both sides (bow and stern),
while it oscillates and propagates at a low average speed.
For a wire with a high transverse anisotropy and in a weak field,
the DW emits mostly stern waves, while the DW distorts itself and
DW center propagates forward like a drill at a relative high speed.
\end{abstract}

\maketitle

\section{Introduction}
Manipulation of magnetic domain wall (DW) in nanostructures draws much
attention for its potential applications \cite{Parkin,Cowburn,Review}
and academic interest as a nonlinear system \cite{Walker, Thiele}.
DW dynamics is governed by the Landau-Lifshitz-Gilbert (LLG) equation.
Spin waves (SWs), collective excitations of spins, can be information
carriers similar as electrons in electronics.
The interplay between SWs and DWs has also received much attention
\cite{magnon,Nowak,YT,entropy,Khodenkov, yanming,wieser,WXS,hubin},
including DW propagation driven by externally generated SWs and SW
generation by a moving DW. Field-driven DW propagation under a
longitudinal field is originated from the energy dissipation that is
compensated by the Zeeman energy released by the DW motion \cite{xrw}.
In the well-known Walker solution \cite{Walker}, the Zeeman energy is
dissipated by phenomenological damping. The DW can even propagate in
a dissipationless wire \cite{wieser,WXS} through emitting SWs and the
Zeeman energy transfers into SW energy \cite{WXS}.
Although the global picture is clear, a microscopic understanding
of how the SWs are emitted in field-driven DW motion is still needed.

In this paper, we numerically study the SW emission by field-driven
DW motion in a wire with two magnetic anisotropy coefficients (biaxial
wire), one for the easy-axis along the wire and the other for
the hard-axis along one of the perpendicular directions of the wire.
In the absence of the Gilbert damping, a transverse DW can propagate
in two distinct modes through emitting SWs. For a broad DW with a low
transverse anisotropy and under a field larger than a critical value,
the DW precesses around the wire while the DW width undergoes a
breathing motion. The DW emits both bow and stern SWs in this case.
This critical field corresponds to the field at which DW precession
frequency is equal to the minimal SW frequency. In contrast, when a DW
is able to shrink to as narrow as the lattice constant due to a high
transverse anisotropy, the DW emits mainly stern SWs and propagates at
an almost constant speed while the spins in the DW are not in a plane.
We further show that in the presence of a small Gilbert damping, above
results are robust despite the decay of SWs.
The properties of the emitted SWs as well as the dependence of DW
propagation speed on applied field and material parameters are obtained.

We consider a classical Heisenberg biaxial chain along z-direction in an
external field $\mathbf{H}$ \cite{wieser,Yanpeng}:
\begin{equation}
\begin{split}
\mathcal{H}=&-J\sum_{n}\mathbf{s}_{n}\cdot\mathbf{s}_{n+1}\\
&-\sum_{n}D_z s_{n,z}^2+\sum_{n}D_x s_{n,x}^2
-\mu_0 \mu_s\mathbf{H}\cdot\sum_{n}\mathbf{s}_{n}.
\end{split}
\label{Heisenberg}%
\end{equation}
$\mathbf{s}_n$ is the unit direction of the spin at lattice site $n$ with
three components $(s_{n,x},s_{n,y},s_{n,z})$. The magnitude of the spin
magnetic moment is $\mu_s=\mu_B S$ with $\mu_B$ the Bohr magneton and $S$
the spin per unit cell. The saturation magnetization is $M_s=\mu_s /a^3$,
where $a$ is the lattice constant. The first term of $H$ is the ferromagnetic
($J>0$) exchange energy. The second and third terms describe easy- and
hard-axis anisotropy energies with coefficients $D_z$, $D_x>0$, and the
last term is the Zeeman energy. The dipolar field is approximately included
in $D_x$ and $D_z$ \cite{wieser,Yanpeng} as the shape anisotropies.

The spin dynamics is governed by the LLG equation \cite{wieser},
\begin{equation}
\frac{\partial \mathbf{s}_n}{\partial t}=-\gamma\mathbf{s}_n\times
\mathbf{H}_{\mathrm{eff},n}+\alpha\mathbf{s}_n\times\frac{\partial
\mathbf{s}_n}{\partial t},  \label{LLG}
\end{equation}
where $\mathbf{H}_{\mathrm{eff},n}=-\partial\mathcal{H}/{\mu_S}\partial
\mathbf{s}_n$ is the effective field. $\gamma$ is the gyromagnetic ratio,
and $\alpha$ is the Gilbert damping. To investigate the DW motion and SW
emission, we numerically solved Eq. \eqref{LLG} with a static head-to-head
transverse DW initially at the wire center ($n=0$ with $n\in[-5000,5000]$).
To avoid SW reflection at both ends, absorbing boundaries are applied on
both sides by assigning a large damping constant near the ends. Different
choices of material parameters gives different DW motion and SW emission mode.
In the simulations below, the time, length, field and energy are in units of
$(\gamma M_s)^{-1}$, $a$, $M_s$ and $a^3 \mu_0 M_s^2$, respectively.

\section{Breathing Motion}

\begin{figure}[!htb]
\begin{center}
\includegraphics[width=8.0cm]{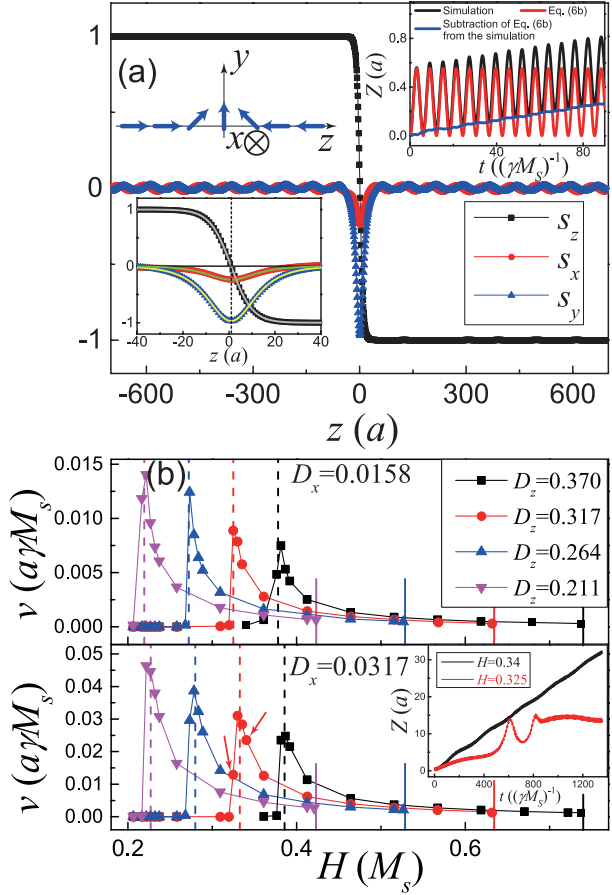}
\end{center}
\caption{(Color online) (a) Schematic diagram of a 1D head-to-head DW
(top-left inset) and the snapshot of $\mathbf{s}$ components at $t=200$
for $\alpha=0$, $J=53.7$, $D_z=0.317$, $D_z/D_x=10$ and $H=0.5$.
Lower-left inset: the $\mathbf{s}$ profile of DW. Symbols are numerical
results, vertical dashed line indicates the DW center position at $z=0.9$,
and solid lines are Walker profile \eqref{Walker2} with $\phi=Ht$ and $Z=0.9$.
Right inset: the simulated motion evolution of DW center (black curve),
the time dependence of DW center by collective coordinate model (red curve)
and their difference (blue curve).
(b) The field dependence of average DW speed $v$ for different $D_z$ and $D_x$.
Vertical dashed and solid lines correspond to critical fields $H_{c1}$ and
$H_{c2}$, respectively. Inset: simulated DW center for fields below and above
$H_{c1}$ (indicated by arrows in the main figure).}
\label{fig1}
\end{figure}
In this section, we focus on a wire with a broad DW ($J/D_z\gg 1$) and a low
transverse anisotropy $D_z/D_x\gg 1$. We first consider the dissipationless
case ($\alpha=0$) with $J=53.7$ (YIG parameter \cite{YIG}), $D_z=0.317$ and
$D_z/D_x=10$. The snapshot of spin profile near DW center at $t=200$ is
shown in Fig. 1(a) when a field of $H=0.5$ is applied at $t=0$.
Initially, the spin profile follows the well-known
Walker solution \cite{Walker,Yanpeng,wieser},
\begin{equation}
\theta_n=2\arctan e^{\frac{n}{\sqrt{J/(2D_z)}}},\qquad \phi_n=\frac{\pi}{2},
\label{Walker1}
\end{equation}
with the DW width $\Delta=\sqrt{J/(2D_z)}$. $\theta_n$, $\phi_n$
are polar angle and azimuthal angle of spin at site $n$. The DW emits SWs
to both sides after the field is applied. The time dependence of the DW
center $Z$, defined as the location where the spin has zero $z$-component,
is shown in the right inset of Fig. 1(a). The DW plane precesses with
frequency $H$ while DW center $Z$ oscillates periodically with frequency $2H$
and moves slowly and simultaneously along the field direction.
The oscillatory DW motion can easily be explained by energy dissipation
theory \cite{xrw} though it is widely understood by the well known
collective coordinate model \cite{suhl,szhang,tatara,tatara1}, in which the
DW is assumed to have a constant azimuthal angle $\phi(t)$ (no twisting).
The polar angle $\theta_n(t)$, with DW width $\Delta{(t)}$, follows the
Walker form
\begin{equation}
\begin{split}
\theta_n(t)=2\arctan \exp\frac{n-Z(t)}{\Delta(t)},\\
\Delta(t)=\sqrt{\frac{J}{2[D_z+D_x \cos^2 \phi(t)]}}.
\end{split}
\label{Walker2}
\end{equation}
The DW dynamics is described by $\phi(t)$ and $Z(t)$,
\begin{equation}
\begin{gathered}
\frac{d\phi(t)}{dt}+\frac{\alpha}{\Delta(t)}\frac{d Z(t)}{dt}=H,\\
-2D_x \sin\phi(t)\cos\phi(t)=\frac{1}{\Delta(t)}\frac{d Z(t)}{dt}-\alpha\frac{d\phi(t)}{dt}.
\end{gathered}
\label{coll}
\end{equation}
When $\alpha=0$, the solution of Eq. \eqref{coll} is
\begin{subequations}
\begin{equation}
\phi=Ht,
\label{solu1}
\end{equation}
\begin{equation}
Z=2\frac{\sqrt{2JD_z}}{H}[\sqrt{1+\frac{D_x}{D_z}}-\sqrt{1+\frac{D_x}{D_z}\cos^2(Ht)}].
\label{solu2}
\end{equation}
\label{colsolu}
\end{subequations}
Obviously, this is an exact solution of Eq. \eqref{LLG} only when $D_x=0$.
It cannot describe the SW emission and the slow DW center propagation.
The lower-left inset of Fig. 1(a) shows the spin profile from simulations
(symbols) and from the collective coordinate model (solid curves) of Eqs. (4)
and (6a) at $t=200$, with $Z=0.9$  instead of $0.29$ by Eq. \eqref{solu2}.
One can see the spin profile near the DW center nevertheless follows Eqs.
\eqref{Walker2} and \eqref{solu1} extremely well.
Therefore, the DW width follows $\Delta{(t)}=\sqrt{J/2[D_z+D_x \cos^2
\phi(t)]}$, a breathing motion with frequency $2H$.
As shown in the right inset of Fig. 1(a), one can obtain the ``pure" DW
center propagation by subtracting Eq. \eqref{colsolu} (red curve) from
numerically obtained $Z(t)$ (black curve). The results are plotted as the
blue curve that is almost linear whose slope gives the average DW speed.

The field dependence of the average DW speed is shown in Fig. 1(b).
Two critical fields $H_{c1}$ and $H_{c2}$ appear.
In a field larger than $H_{c2}=2D_z$, there is no well-defined DW motion
because the domain antiparallel to the applied field becomes unstable
(the total effective field in the right domain is $(H-2D_z)\hat{z}$
that is opposite to the spin direction when $H>2D_z$).
Below $H_{c2}$, the domain is stable and so is the DW.
As the field decreases, the average DW speed, though low, increases
and increases rapidly before reaching $H_{c1}$.
In the inset of Fig. 1(b), the time dependence of DW center position
is shown for various fields below and above $H_{c1}$.
Below $H_{c1}$, the DW propagation and SW emission become irregular,
and the average DW speed is very low. Our discussion focus in the
region of $H_{c1}<H<H_{c2}$.
\begin{figure}[!htb]
\begin{center}
\includegraphics[width=8.0cm]{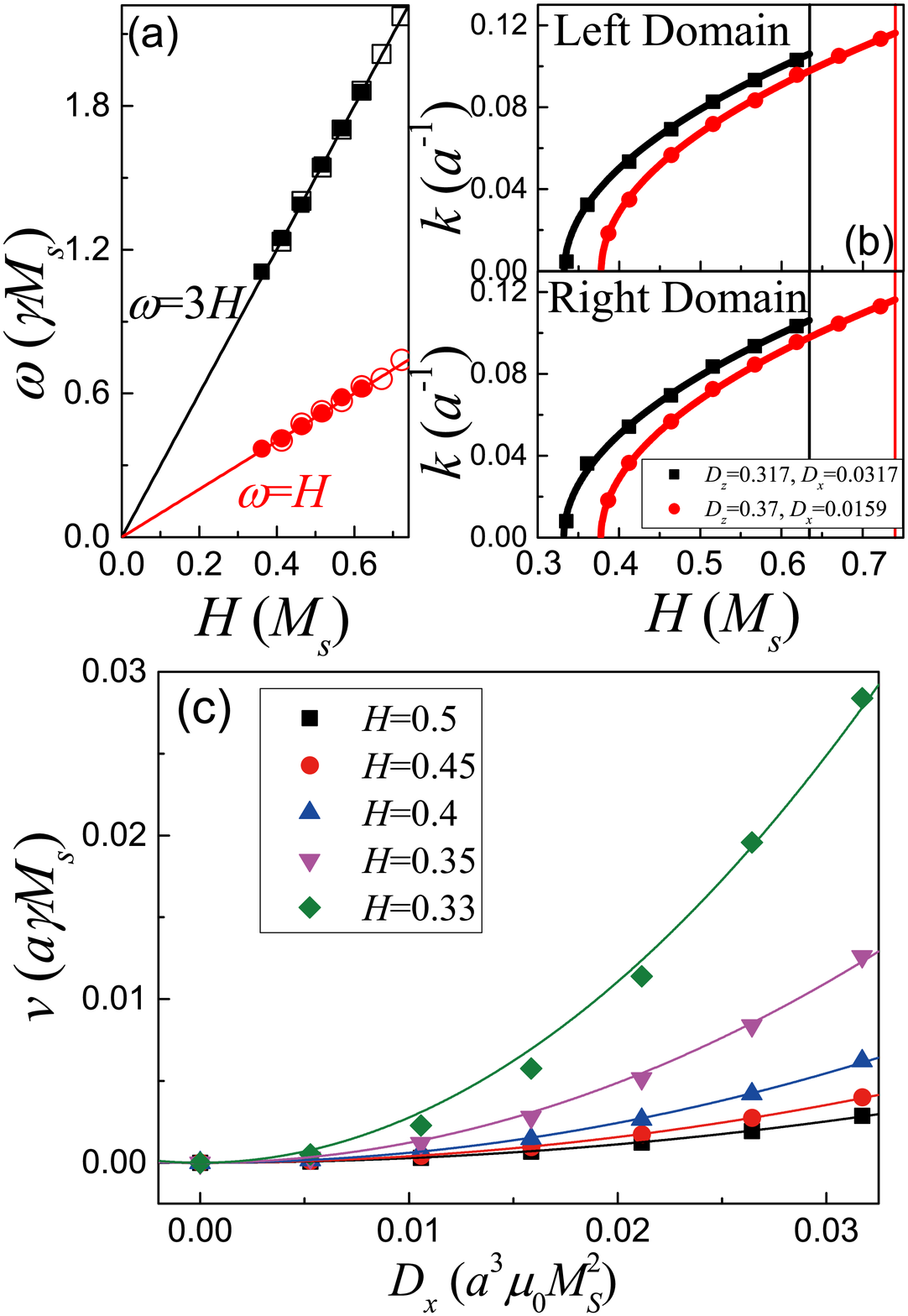}
\end{center}
\caption{(Color online) (a) SW frequency $\omega$ as a function of
applied field $H$. $\omega$ of the left domain is denoted by squares
and that of the right domain by circles. Filled (open) symbols are
results for $D_z=0.317$ and $D_x=0.0317$ ($D_z=0.37$ and $D_x=0.0158$).
(b) SW wave number $k$ as a function of $H$. Upper (lower) part is for
the left (right) domain. Solid curves are analytical results of Eq.
\eqref{dispersion} with $\omega_+=3H$ (upper) and $\omega_{-}=H$ (lower).
Vertical lines indicate critical field $H_{c2}$.
(c) Symbols are $D_x$ dependence of average DW speed for 
$D_z=0.317$ and various $H$.
Solid curves are parabolic fits to the data. }
\label{fig2}
\end{figure}

In order to understand the mechanism of SW emission, we analyze the
wave number $k$ and frequency $\omega$ from the Fourier transforms
of the spatial distributions of $s_i$ (i=x,y,z) at a fixed time
and the time dependence of a spin at a chosen point.
The field dependence of SW frequency is plotted in Fig. 2(a) for a spin
at $n=-100$ (left domain) and $n=100$ (right domain), for several
different $D_x$ and $D_z$. The solid lines are $\omega=3H$ and $\omega=H$.
The frequency in the left domain ($n=-100$) is close to $3H$ and that in
the right domain ($n=100$) is close to $H$, independent of $D_x$ and $D_z$.
The SW has good monochromaticity when $D_x/D_z\ll1$, and the
monochromaticity gradually degrades as $D_x/D_z$ increases.
Following Wieser \textit{et al.} \cite{wieser}, the dispersion relation
of SW can be obtained from Eq. \eqref{LLG} by small fluctuation expansion,
\begin{equation}
\begin{split}
\omega_{\pm}(k)=&\sqrt{2J(1-\cos k)+2D_z\pm H}\\
&\times\sqrt{2J(1-\cos k)+2D_z+2D_x\pm H},
\end{split}
\label{dispersion}
\end{equation}
where ``+" and ``-" are respectively for the left and the right domains.
The wave number $k$ in the left and the right domains can then be uniquely
determined by $3H=\omega_{+}(k)$ and $H=\omega_{-}(k)$, respectively.
This relation is well verified by the numerical results shown in Fig. 2(b).
The upper (lower) part is the field dependence of $k$ for the left (right)
domain. The solid curves are analytical results Eq. \eqref{dispersion}.
All numerical data (symbols) fall on the analytical curves.
The spectrum is gapped with the minimal frequency
$\omega_0=\sqrt{(2D_z\pm H)(2D_z+2D_x\pm H)}$.
According to the early discussion, the DW precesses counter-clockwisely
around $z$-axis at frequency $H$ and DW width breathes at frequency $2H$.
It has already been shown that the breathing of DW width can emit SWs
\cite{WXS} and the generated SWs have the same frequency as that of the
breathing. Thus, in the rotating frame of DW, the DW breathing emits
SWs of frequency $2H$ to both sides. Because the SW in the left (right)
domain is the collective spin procession motion in which spins precess
counter-clockwisely (clockwisely) around the $z$-axis with frequency $2H$,
the spin precession frequency in the left (right) domain is $2H+H=3H$
($2H-H=H$) in the laboratory frame. To emit monochromatic SWs to both
sides of the DW, spin precession frequencies on the left and right domains
should match with the SW spectrum Eq. \eqref{dispersion} that has a gap of
$\sqrt{(2D_z\pm H)(2D_z+2D_x\pm H)}$. Thus $H_{c1}$ should be the larger
$H$ satisfying
$3H=\sqrt{(2D_z+ H)(2D_z+2D_x+H)}$ and $H=\sqrt{(2D_z- H)(2D_z+2D_x-H)}$.
When $D_x\ll D_z$ as used here, we have $H_{c1}\approx D_z+D_x/2$ which
are shown by the vertical dashed lines in Fig. 2(b), and they agree with
the numerical results well. This is called ``breathing motion" because
DW breathing is the origin of SW emission.
SW emission can still occur below $H_{c1}$ but the emitted SWs are
non-monochromatic because individual spin precession cannot be synchronized.
Both the SW pattern and DW motion become complicated and irregular.

The average DW speed $v$ as a function of transverse anisotropy $D_x$ in
a fixed field is shown in Fig. 2(c). When $D_x=0$, Eq. \eqref{colsolu}
is the exact solution, and no DW breathing or SW emission can occur.
The average DW speed is zero. As $D_x$ increases, $v$ increases rapidly.
Numerical data can be fit by a parabolic function $v=C D_x^2$ (solid curves).
The fitting is better for $H$ far above $H_{c1}$, and becomes worse (green
curve) for $H$ close to $H_{c1}$.

The similar simulations were repeated for non-zero damping of
$\alpha=0.001$ (reasonable for permalloy but $1\sim2$ orders of magnitude higher
than that of YIG). The snapshots of spatial distributions of
three spin components at $t=200$ is shown in Fig. 3 with the same
parameters as those in Fig. 1(a) except $\alpha$. The right inset is the
time dependence of DW center position $Z(t)$ from numerical simulations
(black curve) and theoretical prediction from Eq. \eqref{coll} (red curve).
The deviation is due to the SW emission and resulted DW propagation.
The SW pattern is qualitatively the same as that of zero damping
case except the decay of SW amplitude. The nice DW breathing
motion and SW emission exist clearly in the presence of damping.
Because of the energy conservation \cite{xrw,WXS}, the average DW speed
connects directly to the SW emission power density $P$ by $P=2Hv$.
Thus, Figs. 2(b) and (c) give also $H$ and $D_x$ dependences of $P$
with a proper factor of $2H$
\begin{figure}
\begin{center}
\includegraphics[width=8.0cm]{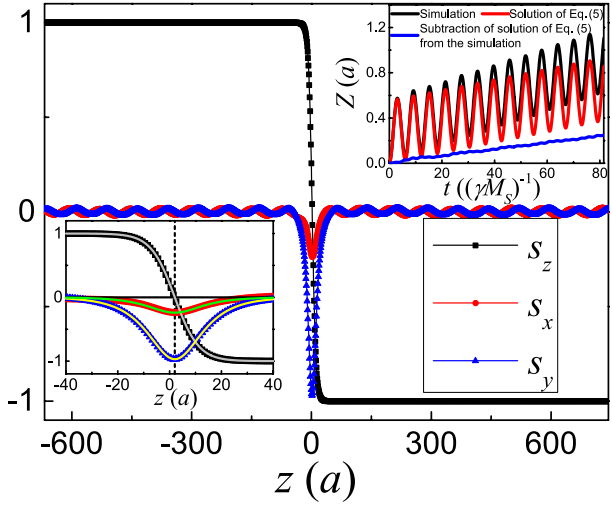}
\end{center}
\caption{(Color online) Snapshot of $\mathbf{s}$ components at $t=200$
for $J=53.7$, $D_z=0.317$, $D_z/D_x=10$, $H=0.5$ and $\alpha=0.001$.
Left inset: $\mathbf{s}$ profile of DW. Symbols are numerical results.
Vertical dashed line indicates the DW center position at $z=2$, and solid
curves are Walker profile \eqref{Walker2} with $\phi=Ht$ and $Z=2$.
Right inset: $Z(t)$ vs $t$ from simulations (black curve) and from
solution of the collective coordinate model Eq. \eqref{coll} (red curve).
The difference (blue curve) shows the extra DW propagation due to SW emission.}
\label{fig3}
\end{figure}

\section{Drilling Motion}

\begin{figure}[!ht]
\begin{center}
\includegraphics[width=8.0cm]{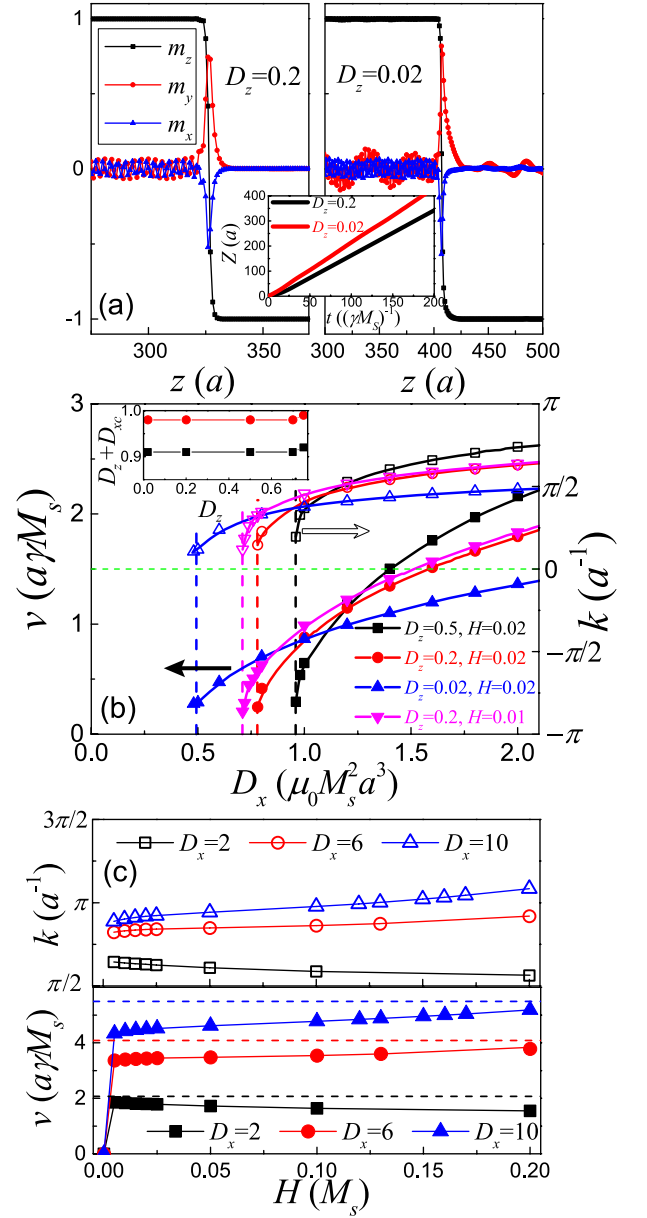}
\end{center}
\caption{(Color online) (a) Snapshot of three spin components at
$t=200$ for $J=2$, $D_x=2$, $\alpha=0$, $H=0.02$ and $D_z=0.2$ (left
panel) or $D_z=0.02$ (right panel). Inset: Time-dependence of the DW
center position for $D_z=0.2$ (black line) or $D_z=0.02$ (red line).
(b) DW speed $v$ (left axis) and SW wave number $k$ in left domain
(right axis) as a function of transverse anisotropy $D_x$.
The dashed lines indicate the smallest $D_x$ for drilling motion.
Inset: $D_z$ dependence of $D_z+D_{xc}$ for $H=0.02$ (red circles) and
$H=0.01$ (black squares). The unit is $\mu_0M_s^2 a^3$.
(c) $k$ (upper panel) and  $v$ (lower panel) as a function of
applied field $H$ for $D_z=0.2$.
The dashed lines indicate the largest speed from Eq. \eqref{coll}.}
\label{fig4}
\end{figure}

Fig. 4(a) is the snapshot of spatial distributions of three spin
components at $t=200$ for $J=2$, $D_x=2$, $\alpha=0$, $H=0.02$
and $D_z=0.2$ (left panel) or $D_z=0.02$ (right panel).
This wire has a strong hard axis ($D_x\gg D_z$).
The wavy patterns away from the DW region are the emitted SWs.
When $D_z=0.2$ only stern waves were observed while there
are both bow and stern waves in the case of $D_z=0.02$.
However, the stern waves are much bigger than the bow waves.
The time dependence of DW center position is shown in the inset.
The linear feature indicates a constant speed of DW propagation,
completely different from the prediction of collective coordinate model.
The azimuthal angle $\phi$ of the spins near the DW center is not a
constant, meaning that the DW is distorted. The distortion comes
from complicated misalignments of spins and their local effective
fields under an external field and a high transverse anisotropy.
Furthermore, $\phi$ angle at DW center does not rotate \cite{WXS}.
The DW emits SWs and moves forward like a drill.

The dependence of DW speed $v$ on transverse anisotropy $D_x$
is shown in Fig. 4(b) (left axis). There is a threshold $D_{xc}$
above which the drilling motion was observed.
$v$ increases rapidly with $D_x$ near but above $D_{xc}$ and the
incremental rate slows down for larger $D_x$.
In order to find out the necessary condition for the drilling motion,
simulations with different fixed $D_z$ and $H$ were carried out.
$D_{xc}$ increases with $H$ for a fixed $D_z$. For a fixed $H$, the value
of $D_z+D_{xc}$ is almost independent on $D_z$, as shown in the inset of
Fig. 4(b). This indicates that the sum of the longitudinal and transverse
anisotropies $D_z+D_{x}$ determines the drilling motion, instead of ratio
$D_x/D_z$ conjectured in previous publications \cite{WXS,wieser}.
A DW undergoes a drilling motion only when $D_z+D_{x}$ is close to $J$.
Since DW width is of order of $\sqrt{\frac{J}{2(D_z+D_x)}}$, this means
that DW drilling motion appears when DW width approaches the lattice constant.
This can be seen in Fig. 4(a) where $s_z$ varies from 1 to -1 within a few
lattice sites. Thus, the motion of the DW can be described as follows:
When a longitudinal field is applied to a static DW, the spins inside
the DW experience a torque and rotate around $z$-axis, resulting in a
finite component along hard axis ($x$-axis) and contraction of DW width.
The strongly disturbed DW starts to emit SWs, and, as a consequence of
energy conservation \cite{xrw}, the DW has to propagate at a constant speed
to release the Zeeman energy to compensate the energy carried away by SWs.

The DW speed $v$ shows a complicated dependence on applied field $H$, as
shown in Fig. 4(c) (lower part). The DW speed $v$ jumps to a high speed
at a very weak field. For a very large $D_x$, $v$ increases slowly with $H$.
However, for a smaller $D_x$, $v$ decreases with $H$.
Although the DW is twisted and the collective coordinate model completely
fails in the drilling motion, the DW speed cannot exceed the largest speed
predicted by Eq. \eqref{coll}, which ($\alpha=0$) is
\begin{equation}
v_m=\sqrt{2J}\sqrt{(D_x+2D_z)-2\sqrt{D_z(D_z+D_x)}},
\end{equation}
(dashed lines in Fig. 4(c)). Interestingly, $v_m$ is also the maximal
velocity of well-known solitons existing at $\alpha=0$ and $H=0$.
A critical field $H_{c}=2D_z$ also exists
in drilling motion for the same reason discussed in breathing motion.
The dependence of wave number $k$ on $D_x$ and $H$ is shown in Figs. 4(b)
(right axis) and 4(c) (upper panel), respectively, with the same trend as
that of $v$. The wave number can be as large as $\pi$, indicating that shot
wave with wavelength comparable to the lattice constant can be emitted.
The dispersion relation Eq. \eqref{dispersion} is verified by the
numerical results.
\begin{figure}
\begin{center}
\includegraphics[width=8.0cm]{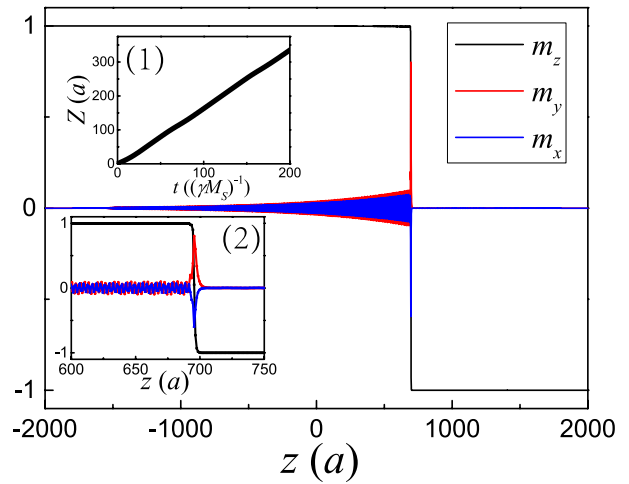}
\end{center}
\caption{(Color online) Snapshot of spatial distributions of spin components
at $t=200$ with $J=2$, $D_z=0.2$, $D_x=2$, $\alpha=0.001$ and $H=0.05$.
Inset (1): The time dependence of DW center position.
Inset (2): Enlarged figure near the DW center.}
\label{fig5}
\end{figure}

Fig. 5 is the snapshot of spatial distributions of $s_x$, $s_y$ and
$s_z$ at  $t=200$ with the same parameters as those in Fig. 4(a) except
$\alpha=0.001$ and $H=0.05$. The Walker
solution \cite{Walker} doesn't exist since $H>H_W\equiv\alpha D_x$.
The DW still propagates at a constant speed while SWs are emitted.
The spatially decaying pattern is due to the non-zero $\alpha$.
The energy dissipation is through both the SW emission and the
Gilbert damping. Clearly, the drilling motion with emission of SWs
exists irrespective of the presence of the Gilbert damping or not.

Our numerical results show that the drilling motion may relate to
strong twisting of DWs, to atomic scale, that may fail the
continuum description of LLG equation \cite{Walker,tatara,tatara1}.
Thus, our discussions do not apply $D_z$ close to or larger than $J$
because the static DW width would be of atomic scale. All initial DW
configurations in our simulations follow the Walker profile well.
A broad DW shrinks a little by a low transverse anisotropy, the DW
undergoes a breathing motion. A high transverse anisotropy distorts the
DW to an atomistic scale, and the drilling motion takes over DW dynamics.
The average DW speed in breathing motion is quite low.
Using YIG parameters $M_s=1.94\times10^5$ A/m, $\gamma=28$ GHz/T and $a=
1.23$ nm, the DW motion shown in Fig. 3 has a speed of 6 cm/s. The total SW
emission power density per unit cross-section area is $P=2vH=2.8$ kW/m$^2$.
DW speed in drilling motion is much higher. The estimated average speed
is 132 m/s for YIG parameters and $D_z=0.1J$, $D_x=0.5J$ and $H=0.02J$.
The total SW emission power is $2.5\times10^2$ kW/m$^2$.
For a nanoscale magnetic wire, the power can be of order of pW which
is applicable in devices such as logic circuits \cite{KLWang}.
However, the anisotropy values used in this work are not realistic, at
least for YIG because the crystalline anisotropy of YIG is about $0.02$
and the hard-axis shape anisotropy of a film is 0.5, which are much small
than $J=53.7$. So in order to experimentally study the drilling motion,
the material has to have high crystalline anisotropy and/or large
saturation magnetization, and relatively weak exchange stiffness.
All of our simulations were done by using \textsc{oommf} \cite{oommf} and
M\textsc{u}M\textsc{ax} \cite{mumax} that agree with each other well.
Although SW emission is discussed in the context of energy dissipation and
DW motion as a consequence of the energy dissipation \cite{WXS,xrw}, there
are many other aspects about the interaction between DWs and SWs.
For example, as discussed by Y. Le Maho \textit{et al.} \cite{maho}, an
effective Gilbert damping constant was defined there due to SW emission.
A modified damping constant can also modify a DW motion. The interaction
between DWs and SWs may lead to a modification of the domain wall width
and an effectively defined mass.

\section{Conclusions}
We numerically investigated the SW emission in field-driven DW motion.
Two modes, breathing motion and drilling motion, were identified.
In strong field and low transverse anisotropy, a broad DW undergoes
a breathing motion. Both bow and stern SWs are emitted by the periodical
breathing of DW width. The main frequencies of the SWs are $3H$ and $H$
in the domains along and opposite to the field, respectively.
There is a lower critical field $H_{c1}$, due to the gap of SW spectrum,
above which monochromatic SW is emitted.
Only long wave can be emitted because applied field is limited by
an upper critical value $H_{c2}$.
The monochromaticity of the SWs degrades when the field decreases and/or
the transverse anisotropy increases. When the sum of longitudinal and
transverse anisotropy energies is of the same order of the exchange energy,
the DW goes into drilling motion when DW plane is greatly distorted.
The DW propagates at a constant high speed, and stern waves are mainly emitted.
The SW wave number and DW speed increase with transverse anisotropy, and
their field dependence is very complicated. The wavelength of the emitted
SW can be very short (close to the lattice constant). Both breathing and
drilling motion are robust against the presence of the Gilbert damping.
The dispersion relation of the emitted SWs follows the theoretical formula.

\section{ACKNOWLEDGEMENTS}
This work was supported by the NSF of China grant (11374249) and Hong Kong
RGC grant (605413).


\end{document}